\documentstyle[12pt]{article}
\input epsf

\newcommand{\be}{\begin{eqnarray}}
\newcommand{\ee}{\end{eqnarray}}

\begin{document}
\title{Gaugino Mass dependence of Electron and Neutron
Electric Dipole Moments}
\author{L. Clavelli\footnote{lclavell@bama.ua.edu}\\
Department of Physics and Astronomy, University of Alabama,\\
Tuscaloosa AL 35487\\ 
\and
T. Gajdosik\footnote{garfield@hephy.oeaw.ac.at} 
\enspace and
W. Majerotto\footnote{majer@hephy.oeaw.ac.at}\\
Institut f\"ur Hochenergiephysik\\ 
der \"Osterreichischen Akademie der Wissenschaften,\\
A-1050, Vienna, Austria\\}
\maketitle
\begin{abstract}
Unless squarks and sleptons
are in the multi-TeV region or above, CP violating
induced electric dipole moments of elementary particles
can pose significant puzzles for supersymmetric
(SUSY) models.  We study
the dipole-moment-inducing one-loop amplitudes as a
function of the fundamental SUSY parameters and show
that these puzzles are removed if there is a
sufficiently large hierarchy between the
gaugino masses and
the scalar mass. We comment on the
experimental status of the low gaugino mass scenario.
\end{abstract}
\renewcommand{\theequation}{\thesection.\arabic{equation}}
\renewcommand{\thesection}{\arabic{section}}
\section{\bf Introduction}
\setcounter{equation}{0}
     It is well known that the standard model by itself does not
provide a sufficiently strong source of CP violation to ensure
baryogenesis assuming an initially matter-antimatter symmetric
universe.  From this point of view, the extra sources of CP violation
natural in a supersymmetric (SUSY) theory are welcome additions to
the theory.  In SUSY the Higgs mixing parameter $\mu$ in the 
superpotential and the soft SUSY breaking Majorana gaugino masses 
and trilinear couplings generically contain CP violating phases.
These can be written
\be
  {\cal L}_{\it soft} = \frac{1}{2} M_a \lambda_a \lambda_a
  - \epsilon_{jk} \left ( 
      h_2^j {\tilde u}_R^\star A_u {\tilde q}_L^k 
    - h_1^j {\tilde e}_R^\star A_\ell {\tilde \ell}_L^k
    - h_1^j {\tilde d}_R^\star A_d {\tilde q}_L^k 
    \right ) + h.c. 
  + \dots
\enspace .
\ee

There is no known mechanism, once SUSY breaking is introduced, to
have the phases in the gaugino masses $M_a$, the Higgs mixing parameter
$\mu$,
or the trilinear couplings A anomalously small and indeed,
from the point of view of baryogenesis, one might hope that these
phases are near maximal.  On the other hand, the precision limits
\cite{bounds}
on electron and neutron electric
dipole moments (EDM's) then pose significant problems
for SUSY:
\be
     |d^e| \leq 2.15 \; 10^{-13} e/GeV  \\
     |d^N| \leq 5.5 \; 10^{-12} e/GeV 
\ee
  
It is known \cite{theory,a1,a2,a3} that, if all
SUSY particles are near some common scale $m_0$,
induced electric dipole moments greatly exceed the
experimental limits unless a) $m_0$ is of order several
TeV, b) the naturally occuring SUSY phases are
extremely small, or c) highly fine-tuned relationships
exist between the SUSY phases and the other parameters
of the theory (masses and coupling constants).
Each of these is problematic for the theory from the point of
view of naturalness, radiative breaking of the electroweak
symmetry and baryogenesis \cite{CQW}.  For a review see
\cite{Cohen}.
A compromise
combining aspects of all three of these solutions is also not
totally satisfactory and
it is still interesting to ask whether there is an alternative
solution which preserves the possibility of SUSY breaking near
the \mbox{100 GeV} scale and possible maximal CP violations in other
processes.

     In the Lagrangian one of the complex fermion mass parameters 
$M_{1,2,3}$ and $\mu$ can be made real by a phase transformation of 
the fermion fields, shifting the phase to the other parameters. 
Thus, it is clear that if the gaugino masses and the A parameters go 
to zero, the theory becomes CP conserving independent of the value of
the $\mu$ parameter.  In this article we investigate how small the
gaugino masses must be to respect the experimental limits on the
dipole moments for low values of the scalar masses.

     Such a hierarchy can be naturally obtained by imposing on
the Lagrangian an approximate R symmetry which then guarantees
vanishing gaugino masses and A parameters at tree level.  This
corresponds to the scenario in which gluino and photino masses
are very light and the other gauginos are in the W,Z mass region.
Although direct searches for light gluinos have been negative up
to now and some experimental counterindications have been
presented, the scenario is still not conclusively ruled out.
The opposite scenario in which gaugino masses are orders of
magnitude above the scalar masses is another possibility which
can solve the dipole moment problem while still leaving squarks
and sleptons at the \mbox{100 GeV} scale.  It is, however, not clear
whether this can be done in a natural way.

     The structure of the paper is as follows.  In section II
we discuss the gluino contributions to the quark electric dipole
moments and present the excluded regions in the $m_0, M_3$ plane
if this contribution is to separately respect the experimental
limits on the Neutron dipole moment with maximum CP violating phase.
In this paper, unless otherwise specified, $m_0$ and $A$ are the
scalar mass and trilinear coupling appropriate to the first generation
only.
In section III, we discuss
the corresponding limits in the $m_0, M_2$ plane assuming the chargino
contribution to the electron EDM is similarly consistent with
experiment.  The results of sections II and III show that the
EDM's go to zero if $M_2$ and $M_3$ tend to zero (or infinity)
for fixed scalar masses.

     In section IV we analyze the more complicated neutralino
contributions and show that these also vanish if the tree level
gaugino masses (together with the A parameter) tend to zero
even though the physical gaugino mass eigenstates are then of order
$M_W^2$.

     In section V we discuss the current viability of the low
gaugino mass scenario and other alternative solutions suggested
by the present work.

\section{\bf Gluino Contributions to Quark Diple Moments}
\setcounter{equation}{0}

     The simplest SUSY contribution to the electric dipole moments
is that of the gluino.  It can be written \cite{theory}

\be
   d_{\tilde g}^q = - \frac{2 e m_q \alpha_s Q_q}{3 \pi ({\tilde m}_1^2
   - {\tilde m}_2^2)} Im \left ( M_3 (\mu \Theta(\beta) - A_q^\star) \right )
      \sum_{k=1}^{2} \frac{(-1)^k}{{\tilde m}_k^2} B(M_3^2/{\tilde m}_k^2)
\ee
Here ${\tilde m}_k$ is the mass of the k'th scalar partner of the quark q,
with
${\tilde m}_1$ being the lighter of the two.  We have neglected terms higher
than
first order in the quark mass $m_q$.
$\Theta (\beta)$ is $(\tan \beta )^{(-2T_3)}$ for a quark of
weak isospin $T_3$. For large $\tan \beta$ the dipole moment of the down
quark
becomes greater, exacerbating the dipole moment problem.  We therefore
assume $1.2<\tan \beta<3$.  The $B$ function is
\be
       B(r) = \frac{1}{2} + r A(r)
\label{eq:B}
\ee
with
\be
      A(r) = \frac{ \ln r}{(1-r)^3} + \frac{3-r}{2 (1-r)^2}
\label{eq:A}
\ee

Note that $r^{(1/2)} B(r)$ has an inversion symmetry.
\be
  r^{1/2} B(r) = r^{-1/2} B(r^{-1})
\ee

It is clear, therefore, that the gluino contribution to the quark
(and consequently to the neutron) electric dipole moment vanishes if
the gluino mass $M_3$ becomes sufficiently small or sufficiently large
for any fixed squark mass.
In actuality, the gluino contribution is not very constraining for
SUSY. In fig. 1, the shape coded points are values in the $M_3-m_0$
plane in which the gluino contribution to the down quark dipole moment
saturates the experimental limit on the neutron dipole moment for values
of $\tan \beta$ between $1.2$ and $3$ and for indicated values of
$\mu$.  The chosen values of $\mu$ are those which are most likely to
allow viable tree level physical chargino and neutralino masses in the low
gaugino mass limit.  Regions in the plane above the arched curve are
consistent with maximum CP violating phases and regions below the curve 
are experimentally excluded.  If the universal scalar mass $m_0$ is above 
the relatively low value of \mbox{350 GeV}, the gluino contribution to 
the quark dipole moments does not exceed the limit on the neutron EDM 
as can be seen from fig. 1.  On the other hand, if the scalar mass $m_0$ 
is in the region of current experimentation ($\approx$ \mbox{100 GeV}), the 
gluino mass must be either below \mbox{1 GeV} or above \mbox{1 TeV} 
assuming, as before, maximal CP violating phases.

\begin{figure}
\centerline{\epsfxsize=3in \epsfysize=3in \epsfbox{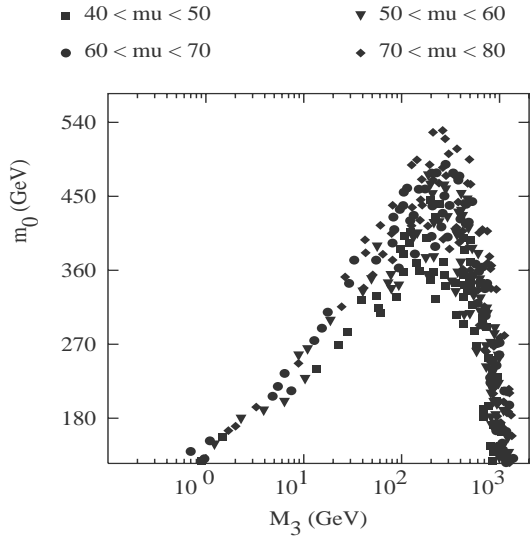}}
\caption{ Minimum values of scalar mass, $m_0$ , as a function of the
SU(3) gaugino mass, $M_3$, for various indicated values of $\mu$
and for $\tan \beta$ between 1.2 and 3.0 assuming the CP violating
phase angle is between 0.95 and 1.0.}
\label{Fig.1}
\end{figure}

\section{\bf Chargino Contributions to the electron EDM}
\setcounter{equation}{0}

     In the presence of CP violating phases, a chargino-sneutrino loop
induces an EDM in the electron.  In minimal SUSY, the chargino is a
mixture of Wino and charged Higgsino.  The tree level mass matrix is

\be
  M_{\chi^{+}} = \left ( \begin{array}{cc} M_2 & \sqrt{2} M_W \sin{\beta} \\
                       \sqrt{2} M_W \cos{\beta} & \mu \end{array} \right )
\ee

As shown in \cite{theory}, the chargino contribution to the electron
EDM is
\be
  d_{\chi^{\pm}}^e = \frac {-e \alpha m_e}{4 \pi \sqrt{2} M_W \cos(\beta)
         {\sin {\theta _W}}^2} \sum_{j=1}^{2} Im \left ( U_{j2} V_{j1}
    \right ) \frac{m_{\chi_j}}{m_{\tilde \nu}^2} A \left ( m_{\chi _j}^2/
     m_{\tilde \nu}^2 \right )
\label{eq:cginocontr}
\ee
The $A$ function, not to be confused with the trilinear A parameter,
is given by Eq.\ref{eq:A}.

$U$ and $V$ are the bi-unitary matrices that relate the tree level
mass matrix, $M_{\chi^{+}}$, to the masses of the chargino eigenstates,
$m_{\chi _j}$.
\be
   \sum_{j} U_{j \alpha} m_{\chi_j}^{2n+1} V_{j \beta} = \left [
       M_{\chi^{+}} \left ( M_{\chi^{+}}^\dagger M_{\chi^{+}} \right ) ^n
       \right ]_{\alpha \beta}
\ee

Thus
\be
  d_{\chi^{\pm}}^e = \frac{-e \alpha m_e}{4 \pi \sqrt{2} M_W \cos{\beta}
     {\sin{\theta_W}}^2 m_{\tilde \nu }^2} Im \left [ M_{\chi^{+}}
     A \left ( M_{\chi^{+}}^\dagger M_{\chi^{+}}/m_{\tilde \nu}^2 \right )
     \right ]_{21}
\ee

By explicit consideration of the general form of the two-by-two
matrices, U and V, one can show that

\be
  d_{\chi^{\pm}}^e = \frac{-e \alpha m_e \tan{\beta} \quad Im(\mu M_2)}{4 \pi {\sin{\theta_W}}^2
      m_{\tilde \nu}^2 (m_{\chi _{+}}^2 - m_{\chi _{-}}^2)}
      \left [ A \left (m_{\chi _{+}}^2/{m_{\tilde \nu}}^2 \right )
    -  A \left (m_{\chi _{-}}^2/{m_{\tilde \nu}}^2 \right )    \right ]
\label{eq:cgino2}
\ee

The mass eigenstates, obtained by diagonalizing
$M_{\chi^{+}}^\dagger M_{\chi^{+}}$, are such that
\be
  m_{\chi_\pm}^2 = M_W^2 + |\mu|^2/2 + |M_2|^2/2 \pm \left [ \left (
   M_W^2 + |\mu|^2/2 + |M_2|^2/2 \right )^2 
   - |M_W^2 \sin{2 \beta} - M_2 \mu |^2 \right ] ^{1/2}
\ee
The properties of the $A$ function are such that the chargino contribution
to the electron EDM vanishes as $M_2$ goes to zero or infinity for any
fixed mass of the scalar neutrino although this behavior is not obvious
from eq.\ref{eq:cginocontr} since for $M_2$ going to zero the chargino
masses are in the $W$ mass region.  Thus, assuming maximum CP violating
phase, the chargino contribution to the electron EDM, by itself, satisfies
the experimental bound for values of the scalar mass $m_0$ above the
curves of fig. 2.  The chargino contribution is seen to be quite
restrictive for the theory.  For scalar masses in the region of current
experimentation ($\approx$ \mbox{100 GeV}), five orders of magnitude in the 
$SU(2)$ gaugino mass are experimentally excluded for maximum phase. 
For $M_2$ in the
$\mbox{100 GeV}$ region, as is often assumed, the scalar mass would have to be
above \mbox{1 TeV} leading to severe problems for the hypothesis of radiative
breaking of the electroweak symmetry.\\

\begin{figure}
\centerline{\epsfxsize=3in \epsfysize=3in \epsfbox{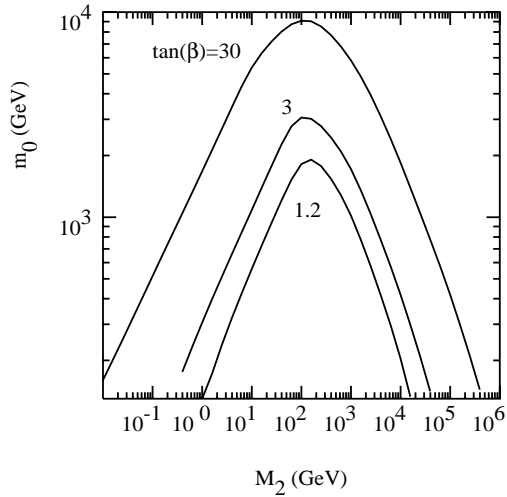}}
\caption{ Minimum values of scalar mass, $m_0$ , as a function of the
SU(2) gaugino mass, $M_2$, for various indicated values of $\tan \beta$.
Here $\mu=50 \; GeV$ and the CP violating phase between $\mu$ and $M_2^\star$
is taken to be $ \phi = 45 \deg $.}
\label{Fig.2}
\end{figure}

With $M_2$ sufficiently small (or sufficiently large) to satisfy the
dipole moment bounds,  
gaugino mass universality would exclude a wider range of
gluino masses than could be directly excluded by the gluino contribution
to the neutron EDM considered in the previous section.

     If the chargino contribution to the electron EDM is, by itself,
consistent with the experimental limit, it will pose no problem for
the neutron EDM since, even though the fermion mass is then an order
of magnitude greater, the experimental limit is almost two orders of
magnitude weaker.

\section{\bf Neutralino Contributions to the electron EDM}
\setcounter{equation}{0}

     Finally we turn to the neutralino contribution to the electron EDM.
In this case we cannot, in general, find a closed form expression
analogous to eq.\ref{eq:cgino2} since the neutralino system is a mixture
of four states.  With basis states ${\tilde \gamma}, {\tilde Z},
{\tilde H}_1, {\tilde H}_2$, the
 tree level mass matrix is given by

\be
M_{\chi^0} =\left (\begin{array}{cccc} M_2 s^2 + M_1 c^2 \quad & (M_2 - M_1)sc \quad & 0  \quad &  0 \quad \\
     (M_2-M_1)sc  \quad & M_2 c^2 + M_1 s^2  \quad & M_Z  \quad & 0  \quad \\
     0  \quad & M_Z  \quad & \mu \sin{2 \beta}  \quad & - \mu \cos{2 \beta}  \quad \\
     0  \quad & 0  \quad & - \mu \cos{2 \beta}  \quad & - \mu \sin{2 \beta}  \quad  \end{array}\right )
\ee
Here $s$ and $c$ are the $\sin$ and $\cos$ respectively of the weak angle.
With $\mu, M_1$ and $M_2$ in general complex, the matrix is non-Hermitian and
the mass eigenstates are found by diagonalizing $M_{\chi^0}^\dagger M_{\chi
^0}$.

     In \cite{theory}, the contribution to the electron EDM is given
in terms of the physical neutralino masses, $m_{\chi_k^0}$,
 and mixing matrices, $N_{jk}$.  Neglecting
terms higher order in the electron mass this can
be written
\be
   d_{\chi^0}^e = \frac{-Q \alpha m_e}{8 \pi s^2 c^2} \sum_{m=1}^{2}
    \frac{1}{{\tilde m}_m^2} \sum_{i=1}^{2} \sum_{j=1}^{4} \sum_{k=1}^{4}
    Im \left [ N_{ik}^\star N_{jk}^\star \gamma_{ij} \right ]
    m_{\chi_k ^0} B \left ( m_{\chi_k^0}^2/{{\tilde m}_m^2} \right )
\label{eq:nlinocontr}
\ee
     Here ${\tilde m}_m$ is the mass of the mth selectron and the coefficients
$\gamma _{ij}$ are
\be
 \gamma_{ij} = a_j \left ( b_i (-1)^m \left ( \frac{\mu
\Theta(\beta) - A^\star}{{\tilde m}_1^2 - {\tilde m}_2^2}(\delta_{j1}
 + \delta_{j2}) + (\delta_{j3}+\delta{j4})/M_Z \right ) + \delta_{i2}
 \delta_{m1}(\delta_{j3}+ \delta_{j4})/M_Z \right )
\label{eq:gamma}
\ee
The coefficients $a_i$ and $b_i$ are given by
\be
\begin{array}{rl}
   a_1  = & - Q \sin(2 \theta_W) \\
   a_2 =  & -1 + 2 Q {\sin(2 \theta_W)}^2\\
   a_3  = & 2 T_3   \\ 
   a_4  = & - \Theta(\beta)  \\ 
   b_i =  & - a_i + \delta_{i2}
\end{array}
\ee
For the electron, $T_3=-1/2$ and $\Theta(\beta)=\tan \beta$.
The unitary matrix $N$ relates the neutralino mass matrix,
$M_{\chi^0}$  to the
physical masses
\be
  \sum_{k=1}^{4} N_{\alpha k}^\star m_k^{2n+1} N_{\beta k}^\star
    = \left [ M_{\chi^0} \left ( M_{\chi^0}^\dagger M_{\chi^0} \right )
      ^n \right ]_{\alpha \beta}
\ee
We can thus write the relation between the electron EDM and the
tree level mass matrix
\be
  d_{\chi^0}^e = \frac{-Q \alpha m_e}{8 \pi s^2 c^2} \sum_{i=1}^{2}
   \sum_{j=1}^{4} Im \gamma_{ij} \left [ M_{\chi^0} B \left(
   M_{\chi^0}^\dagger M_{\chi^0}/{{\tilde m}_m^2} \right ) \right ]_{ij}
\label{eq:nlino2}
\ee
     It is now easy to show that the neutralino contribution to the
EDM vanishes as $M_1, M_2$, and the trilinear coupling $A$ go to zero.
It is clear from eq.\ref{eq:nlino2} that the EDM is proportional to elements
from the first two rows of the neutralino mass matrix $M_{\chi^0}$.
Of these only
the $23$ element survives as $M_1$ and $M_2$ go to zero.
In the limit of vanishing
gaugino masses the dipole moment, therefore, becomes

\be
    d_{\chi^0}^e = \frac{-Q \alpha m_e M_Z}{8 \pi s^2 c^2} \sum_{m=1}^{2}
    \frac{1}{{\tilde m}_m^2} \sum_{j=1}^{4} Im \left [ \gamma_{2j}
    B \left ( M_{\chi^0}^\dagger M_{\chi^0}/{{\tilde m}_m^2} \right )_{3j}
    \right ]
\ee

  $\gamma_{2j}$
is real for $j=3$ or $4$.  Since $B$ is Hermitian, $B_{33}$ is real.  As $M_1$
and $M_2$ go to zero, $(M_{\chi^0})_{k1} \rightarrow 0$ and therefore $B_{31}
\rightarrow 0$.  The remaining terms in the sum on j are then

\be
   d_{\chi^0}^e = \frac{-Q \alpha m_e M_Z}{8 \pi s^2 c^2} \sum_{m=1}^{2}
   \frac{1}{{\tilde m}_m^2} Im \left [ \gamma_{22} B \left ( M_{\chi^0}
   ^\dagger M_{\chi^0} /{{\tilde m}_m^2} \right )_{32} 
   + \gamma_{24} B \left ( M_{\chi^0}
   ^\dagger M_{\chi^0} /{{\tilde m}_m^2} \right ) _{34} \right ]
\ee
With $M_1$ and $M_2$ going to zero,
the only off-diagonal terms in $M_{\chi^0}^\dagger M_{\chi^0}$  are
\be
   (M_{\chi^0}^\dagger M_{\chi^0})_{32} = M_Z \mu^\star \sin(2 \beta)\\
   (M_{\chi^0}^\dagger M_{\chi^0})_{24} = - M_Z \mu \cos(2 \beta)
\ee

together with their conjugate elements. Thus
\be
 B_{32} = (M_{\chi^0}^\dagger M_{\chi^0})_{32} \cdot real = \mu^\star \cdot
   real \\
 B_{34} = (M_{\chi^0}^\dagger M_{\chi^0})_{32} (M_{\chi^0}^\dagger
   M_{\chi^0})_{24} \cdot real = real
\ee

Using the explicit form of $\gamma_{ij}$,
it is then clear that the dipole moment
vanishes for vanishing gaugino masses, $M_1$ and $M_2$, providing the $A$
parameter
also vanishes (or has opposite phase to $\mu$).  The gaugino masses and
the $A$ parameter vanish in a technically natural way if one imposes a
continuous $R$ symmetry on the Lagrangian although we must still discuss
in the next section
whether this is a phenomenologically viable possibility.

     To numerically evaluate eq.\ref{eq:nlino2} we expand the $B$ functions
in inverse powers of the selectron masses, ${\tilde m}_m$.  For simplicity we
put $M_2 =0$ since this gaugino mass is strongly constrained by the chargino
contribution. We keep terms up to order ${\tilde m}_m^{-6}$.

If we consider the matrix argument
\be
 r =  M_{\chi^0}^\dagger M_{\chi^0}/{\tilde m}^2
\ee

one can show that

\be
  A(r)_{jk} = \delta_{jk} A(r_{kk}) -(1-\delta_{jk}) r_{jk}
\frac{A(r_{jj})-A(r_{kk})}
    {r_{jj}-r_{kk}} + {\cal O}({\tilde m}^{-4})
\label{eq:Amatrix}
\ee

The result,
which of course is only valid for small $M_1/{\tilde m}_m$, is shown in Fig. 
3.  Over the region plotted in Fig. 3, the $(1-\delta_{jk})$ terms in
Eq.\ref{eq:Amatrix} gives a small (${\cal O}(10\%)$) contribution.
We expect however that, using the inversion symmetry of the B function,
one should be able to show that the dipole moment contribution 
will also become small for large gaugino masses
at fixed scalar mass.  We have checked numerically that this decoupling
holds.

\begin{figure}
\centerline{\epsfxsize=3in \epsfysize=3in \epsfbox{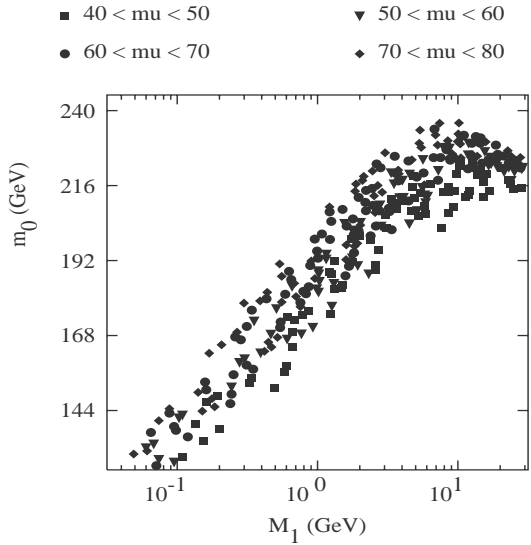}}
\caption{ Minimum values of scalar mass, $m_0$ , as a function of the
U(1) gaugino mass, $M_1$, for various indicated values of $\mu$
and for $\tan \beta$ between 1.2 and 3.0 assuming the CP violating
phase angle is between 0.95 and 1.0.}
\label{Fig.3}
\end{figure}

     As in the case of the gluino contribution, the neutralino contribution
is sufficiently small once the scalar mass is above several hundred GeV.
However, if the squarks and sleptons are in the region of current
experimentation near \mbox{100 GeV}, very small (or perhaps very large) gaugino
masses are required if the CP violating phase is near maximum and there are
no special cancellations.

\section{\bf Status of the low gaugino mass scenario}
\setcounter{equation}{0}

     In the supergravity mediated SUSY breaking scheme, the three
gaugino masses, $M_1, M_2,$ and $M_3$ are expected to be equal at the
unification scale and approximately proportional to the corresponding
gauge couplings at low energies.  However, in more general schemes,
the gaugino masses could widely differ.
Since the early days of SUSY there has been much discussion of the
aesthetics and viability of a light gluino ($M_3 \approx 0$)
\cite{early}.
The primary indication
of a light gluino would be an anomalously slow running of the strong coupling
constant.  In the last decade there were in fact several indications that
this was, in fact, the case \cite{LC92,Hebbeker,CCY,JK,BB}.
More recent re-analyses of
deep-inelastic data and lattice gauge calculations, together with
the widespread feeling that the $\tau$ decay width is the best low energy
measure of $\alpha_s$, have significantly weakened
the case for an anomalously slow running.
There
remain, however, persistent anomalies especially in the quarkonia decay
rates \cite{CC}
requiring an assumption of extremely large and negative relativistic
corrections to the wave functions to obtain consistency with the standard
model.  LEP measurements of the running of $\alpha_s$ in the LEP II region
seem, also, to be consistent with the standard model.  However, in view of the
relatively poor statistics above the $Z$ together with systematic problems
associated with radiative return events, $W$ pair production etc. and in
view of the anomalously large jet production rates observed at Fermilab
and perhaps HERA at high scales, the
case for a standard model running in the high energy region cannot be
considered absolutely settled.  Several features of the Fermilab jet data
and top quark candidates
seem, in fact, in line with what one would expect in the light gluino
scenario \cite{CT1,CG} with, perhaps valence squarks in the \mbox{100 GeV}
region.

     Counterindications to a very light gluino have been published coming
from the four-jet angular distributions \cite{Aleph4j}, and from direct
searches \cite{KTeV,E761}.  The four-jet results, which
suggest a gluino mass of at least $6.3$ GeV, are, however, subject
to criticism \cite{Farrar4j}, and it is possible that
systematic errors related to monte-carlo dependence might be
underestimated.  The negative result from $E761$ could be consistent
with a light gluino if $qqq{\tilde g}$ and $q {\overline q}{\tilde g}$
bound states are not formed for the same reason that no strong candidates
exist for the hybrid states of quarks and gluons $qqqg$ and $q {\overline q}
g$.  Presumably, the QCD potential is too repulsive in the requisite
color octet sub-states.  Similarly the KTeV result would not strongly
impact the light gluino hypothesis if the $R^0(g{\tilde g})$ state is
too long lived as might occur if the photino were not sufficiently lighter
in mass or in some gauge mediated SUSY breaking schemes.

    If the gluino is light (below 10 GeV) it is expected that the charginos
and neutralinos would have dominantly non-leptonic decay modes into
quark-antiquark-gluino.  Thus the traditional signatures for these particles,
isolated leptons and/or significant missing energy, would be invalidated.
A dedicated search involving hadronic decay channels would then be required.
As yet only the OPAL experiment has published the results of such a dedicated
search \cite{OPAL}.  The graph in this reference constraining the chargino
mass is, however, fully consistent with vanishing $M_1$ and $M_2$.  The
resulting parameters, $\mu$ and $\tan \beta$, however, are said to be
inconsistent with the neutralino search if $M_1$ is zero.
    The OPAL work suggests that, assuming a light gluino decay of the
charginos and neutralinos, the origin in the $M_1-M_2$ plane is excluded.
Even this result is in question since they assumed a $100 \%$ hadronic
decay which is unlikely even if the hadronic modes dominate.  An extension
of the OPAL analysis to exclude a finite region in the $M_1-M_2$ plane is
highly desirable.  With $M_1=M_2=0$, the lightest chargino should be
between $50$ and $70$ GeV at tree level but this mass might be increased
somewhat by one-loop corrections.  It would be useful if the LEP experiments
other than OPAL would also publish results on hadronic events in the
$100$ to $180$ GeV region.  Since this overlaps the $W$ pair production
region, it would be important to know whether there is sufficient
flexibility in the Monte Carlos to allow the existence of such charginos
in addition to whether the data is consistent with the standard model.

We would also like to comment on the value of $\tan\beta$, which in
our scenario is $1.2 < \tan\beta < 2$. It strongly determines the
Higgs mass.  In our scenario one would get within the MSSM
a mass of $h^{0}$ between
61.3~GeV (for $\tan\beta = 1.2$) and 77.3~GeV (for $\tan\beta = 2$) for
$m_{A} = 200$~GeV, and the SUSY breaking parameters $A = 0$, $M_{Q_{3}} =
500$~GeV, $M_{U_{3}} = 450$~GeV, $M_{D_{3}} = 500$~GeV.
(The mass scale of the sfermions of the first and second generation is
unimportant). The present experimental
bound on the mass of $h^{0}$, $m_{h} \ge 110$~GeV for $\tan\beta < 3$,
however, strongly relies on the dominance of the decay $h^{0}
\rightarrow b\bar{b}$. In our case the branching ratio of this decay
can be substantially reduced due to the possible decay modes $h^{0}
\rightarrow \tilde{g} \tilde{g}$ \cite{Djouadi} and $h^{0} \rightarrow
\tilde{\chi}^{0}_{2} \tilde{\chi}^{0}_{2}$ with $\tilde{\chi}^{0}_{2}
\rightarrow q\bar{q} \tilde{g}$.  A flavour independent search for a
neutral scalar Higgs was performed by OPAL\cite{2opal}.  According to
this study, a Higgs mass $<$ 66.2~GeV, is excluded for $\sin^{2} (\beta
- \alpha ) \geq 0.5$, where $g \sin (\beta - \alpha )/\cos\theta_{W}$
is the coupling to $Z^{0}$.
Hence a Higgs mass corresponding to our scenario is still possible
especially if the coupling to the $Z^0$ is suppressed.
In this context we would also like to
mention the next--to--minimal supersymmetric extension of the Standard
Model with a gauge singlet superfield added to the Higgs sector
\cite{Ellwanger}. In this model the lightest Higgs boson can
have a weak coupling to the $Z^{0}$ and
would therefore be hardly visible in $e^{+}e^{-} \rightarrow Zh^{0}$.
In such a model the maximum mass of the next to lightest neutral Higgs can
be made to satisfy the current experimental limits with low
values of $\tan{\beta}$ and scalar mass $m_0$.

\section{\bf Conclusions}
\setcounter{equation}{0}

     Detailed investigation of the dependence of the induced electric dipole
moments of the electron and neutron on the SUSY breaking parameters indicates
that the current experimental limits might provide useful hints toward the
structure of the SUSY breaking Lagrangian.  If the gaugino mass parameters
are sufficiently small the squarks and
sleptons might still be in the low energy region near \mbox{100 GeV} and maximum
SUSY CP violating phases are then still possible.
The dipole moment amplitudes then suggest also a vanishing
A parameter and a low $\tan \beta$.  Critical tests of this scenario are
possible at LEP II and in the next run at Fermilab.
The alternative of having one or more gaugino mass parameters very large
with the others small together 
with a small squark mass is also interesting and will be further 
explored in a
future work.

    This work was supported in part by the US Department of Energy under
grant no. DE-FG02-96ER-40967 and by the "Fonds zur F\"orderung der
Wissenschaftlichen Forschung" of Austria, project no P13139-PHY.
LC wishes to thank
the University of Vienna, the Institute of High Energy Physics
of the Austrian Academy of Sciences, and the University of Bonn for
hospitality during the period of this research. We also thank 
H. Eberl and S. Kraml for numerical checks of the Higgs masses.

\end{document}